\documentclass[letterpaper]{JHEP}
\usepackage{epsfig}
\newcommand\fverb{\setbox\pippobox=\hbox\bgroup\verb}
\newcommand\fverbdo{\egroup\medskip\noindent%
			\fbox{\unhbox\pippobox}\ }
\newcommand\fverbit{\egroup\item[\fbox{\unhbox\pippobox}]}
\newbox\pippobox
\title{Essential and inessential features of Hawking radiation}
\author{Matt Visser\thanks{Research supported by the US DOE.}\\
	Physics Department, Washington University in Saint Louis, 
        MO 63130-4899, USA\\
	E-mail: \email{visser@kiwi.wustl.edu}}
\preprint{\hepth{0106111}}	
\abstract{
There are numerous derivations of the Hawking effect available in the
literature.  They emphasise different features of the process, and
sometimes make markedly different physical assumptions.  This article
presents a ``minimalist'' argument, and strips the derivation of as
much excess baggage as possible.  All that is really necessary is
quantum physics plus a slowly evolving future apparent horizon
({\emph{not}} an event horizon).  In particular, neither the Einstein
equations nor Bekenstein entropy are necessary (nor even useful) in
deriving Hawking radiation.}
\dedicated{\rm 13 June 2001; \LaTeX-ed \today}

\begin{document} 
\section{Introduction}
\def\d{{\mathrm{d}}}
\def\implies{\Rightarrow}
\def\Painleve{Painlev\'e}
\def\Barcelo{Barcel\'o}
\def\etal{{\sl et al}}

Hawking radiation from black holes is a semiclassical quantum effect
that has now been with us for some 27 years~\cite{Hawking}, and whose
theoretical importance is difficult to exaggerate.  Over the decades,
the Hawking effect has accreted a quite considerable mythology.
Perhaps the two most pernicious myths attached to this effect are:
\begin{itemize}
\item
``{\sl Hawking radiation has something to do with gravity}'', 
and,
\item
``{\sl Hawking radiation automatically implies Bekenstein entropy}''.
\end{itemize}
These myths were engendered by two historical accidents: (1) the
Hawking effect was first encountered within the context of general
relativity, and (2) the fact that it was discovered shortly after the
notion of Bekenstein entropy (geometric entropy) had been
formulated~\cite{Bekenstein,Laws}.

Though the Hawking effect was at first partly motivated by the need
for a consistent thermodynamic interpretation for the notion of
Bekenstein entropy, it was rapidly appreciated that the Hawking effect
is much more primitive and fundamental~\cite{Unruh-effect} --- in
particular the Hawking effect continues to make sense even in
situations where geometric entropy and even gravity itself are simply
not relevant~\cite{without}. This observation underlies much of the
current interest in ``analog models of/for general
relativity''~\cite{Unruh,Visser}; there is now a realistic possibility
for experimental detection of the Hawking effect in condensed-matter
analog systems using current or planned
technology~\cite{Garay,Leonhardt,SPS}, with ``effective metrics'' and
``black holes'' that have nothing to do with gravity
itself~\cite{Normal-modes}. In view of this situation, in this current
article I will attempt to isolate an irreducible minimum of physical
assumptions needed for the Hawking effect to arise.

Since the literature is vast, I will not be able to do justice to all
known derivations of the effect --- some key derivations are those due
to Hawking himself~\cite{Hawking}, the Hartle--Hawking approach using
analytic continuation of the propagator across the event horizon of an
eternal black hole~\cite{Hartle-Hawking}, and the Gibbons--Hawking
approach using Euclideanization (Wick
rotation)~\cite{Gibbons-Hawking}.  Extremely useful general surveys
are provided in~\cite{IJMPD} and~\cite{Primer}.

A particularly relevant discussion is the early work of Damour and
Ruffini~\cite{Damour}, who emphasise the behaviour of the ``outgoing''
modes as one crosses the horizon. Though their presentation is given
in terms of the Kerr--Newman geometry it is easy to verify that the
specifics of the geometry enter {\emph{only}} in the evaluation of the
surface gravity of the future Killing horizon, and that the route from
surface gravity to Hawking radiation does not depend on either the
Einstein equations or even the underlying physical mechanism leading
to the existence of the metric.

In a slightly different vein, the discussion of Parker~\cite{Parker}
particularly emphasises the relationship with particle production from
a dynamical vacuum state, while Gerlach~\cite{Gerlach},
Grove~\cite{Grove}, Hu~\cite{Bei-Lok}, and Brout and
Parentani~\cite{IJMPD} emphasise in varying degree the near-universal
role of the exponential stretching associated with many types of
horizon.

More recently the contributions of Massar and
Parentani~\cite{Parentani}, {Parikh} and {Wilczek}~\cite{Wilczek}
Padmanabhan~\etal~\cite{Padmanabhan-et-al}, and
Schutzhold~\cite{Schutzhold} should be noted.  They emphasise, in
slightly different forms, the analyticity properties of the modes and
what is effectively an imaginary contribution to the action localized
at the horizon, an observation that can be traced back to the work of
Damour and Ruffini~\cite{Damour}.

It is also important to realise that the Hawking radiation effect is
independent of whatever cutoff you introduce to the high-frequency
physics --- this is one of the theoretical reasons for interest in
analog models, because for acoustic black holes you have an explicit
model for the high frequency cutoff in terms of atomic
physics~\cite{Unruh2,Jacobson2,Parentani2}. I will not have anything
specific to say about this particular issue in the present paper.

Based on the results extracted from 27 years of research, as verified
and made more explicit by the recent interest in analog models, it is
clear that the basic physics we should be aiming for is:
\begin{itemize}
\item
``{\sl Hawking radiation is kinematics}'',
\item
``{\sl Bekenstein entropy is geometrodynamics}''.
\end{itemize}
That is: Hawking radiation is a purely kinematic effect that depends
only on the existence of a Lorentzian metric (with no particular
prejudice as to how this metric arises) and some sort of horizon;
Hawking radiation does not depend on the validity of the Einstein
equations (as may most quickly be verified by looking at Hawking's
original derivation~\cite{Hawking} and verifying that the Einstein
equations are nowhere used nor needed). In contrast, Bekenstein's
geometric entropy associated with the area of the event horizon is an
intrinsically geometrodynamic effect, wrapped up with the validity of
the Einstein equations: Entropy equals one quarter the area (plus
perturbative corrections) if and only if the Einstein equations are
valid (plus perturbative corrections)~\cite{without}. The connection
with the Einstein equations arises because when integrating the first
law to evaluate the Bekenstein entropy you need to use the
relationship between total mass of the black hole and its surface
gravity (and hence Hawking temperature), and it is in this
relationship between surface gravity and mass-energy that the Einstein
equations enter.

I shall also distinguish the notion of ``apparent horizon'' from that
of the ``event horizon'' (absolute horizon) and demonstrate that the
existence of a locally definable apparent horizon is quite sufficient
for obtaining the Hawking effect. (Remember that to define the event
horizon you need to know the entire history of the spacetime out to
the infinite future; you should be a little alarmed if the question of
whether or not a black hole is radiating {\emph{now}} depends on what
it is doing in the infinite future.)

The general theme of the analysis will be to do as much as possible
with the {eikonal} approximation (even {WKB} is mild overkill), and
look for {generic} features in the modes at/near the {apparent}
horizon. Specifically, we will look for a {Boltzmann} factor.  Note
that a good derivation should {\em not} depend on either {grey-body
factors} or a {past horizon}. Grey-body factors are not fundamental
--- they are simply transmission coefficients giving the probability
that modes which escape from the horizon make it out to null
infinity. Past horizons are specific to eternal black holes and simply
not relevant for astrophysical black holes. Similarly because the
entropy-area law is tied to the validity of the Einstein equations we
shall seek to avoid any appeal to this property. Since much of what I
have to say is well-known to experts in the field (at least within the
general relativity community) I will place a premium on clarity and
simplicity.

\section{Metric: \Painleve--Gullstrand form}

In general relativity any {spherically symmetric geometry}, {static or
not}, can by suitable choice of coordinates {locally} be put in the
form
\begin{equation}
\d s^2 = - c(r,t)^2 \;\d t^2 + (\d r - v(r,t) \;\d t)^2    
+ r^2 [ \d\theta^2 + \sin^2\theta \; \d\phi^2 ].
\end{equation}
Equivalently
\begin{equation}
\d s^2 = - [c(r,t)^2 - v(r,t)^2]\; \d t^2 - 2 v(r,t) \;\d r \;\d t 
+ \d r^2 
       + r^2 [\d\theta^2 + \sin^2\theta \; \d\phi^2 ].
\end{equation}
These are so-called \Painleve--Gullstrand coordinates, a relatively
obscure coordinate system currently enjoying a resurgence.
(See~\cite{Poisson} for a geometric discussion.) A nice feature of
these coordinates is that the metric is nonsingular at the
horizon. Though the basic physics is coordinate independent, we shall
see that these coordinates simplify computations considerably.

In the acoustic geometries associated with sound propagation in a
flowing fluid it is often convenient to not set up coordinates this
precise manner but rather to use the natural coordinates inherited
from the background Minkowski spacetime. Doing so results in a minor
modification of the \Painleve--Gullstrand metric
\begin{equation}
\d s^2 = \left[{\rho(r,t)\over c(r,t)}\right]
\left\{ - c(r,t)^2 \;\d t^2 + (\d r - v(r,t) \;\d t)^2    
+ r^2 [ \d\theta^2 + \sin^2\theta \; \d\phi^2 ]\right\},
\end{equation}
where $ c(r,t)$ is now the speed of sound, $ v(r,t)$ is the radial
velocity of the fluid, and $\rho(r,t)$ is its density. Phonons are
then massless minimally coupled scalar fields in this acoustic
geometry~\cite{Unruh,Visser}. Fortunately the conformal factor does
not affect the surface gravity and does not affect the Hawking
effect~\cite{Jacobson-Kang}; for simplicity I shall simply suppress
the conformal factor, it can easily be reinstated if desired.

In matrix form the \Painleve--Gullstrand metric takes the quasi--{ADM}
form
\begin{equation}
g_{\mu\nu}(t,\vec x) 
\equiv
\left[ \matrix{-(c^2-v^2)&\vdots&-v \;\hat r_j\cr
               \cdots\cdots\cdots\cdots&\cdot&\cdots\cdots\cr
               -v \;\hat r_i&\vdots& \delta_{ij}\cr } \right].
\end{equation}
The inverse metric is
\begin{equation}
g^{\mu\nu}(t,\vec x) \equiv 
{1\over c^2}
\left[ \matrix{-1&\vdots&-v \;\hat r^j\cr
               \cdots\cdots&\cdot&\cdots\cdots\cdots\cdots\cr
               -v \; \hat r^i&\vdots&(c^2 \; \delta^{ij} - v^2 \; \hat r^i \; \hat r^j )\cr } 
\right].               
\end{equation}
Because of spherical symmetry finding the apparent horizon is
particularly easy: it is located at $c(r,t) = |v(r,t)|$.  It is easy
to see that the metric is {nonsingular} at the apparent horizon, with
$\det(g_{\mu\nu}) = - c(r,t)^2$.  To get a future apparent horizon,
corresponding to an astrophysical black hole, we need $v<0$.  I make
no claims as to the location or even existence of any event horizon.

[Since for static black holes the event horizon is coincident with the
apparent horizon, the general feeling is that for quasi-stationary
black holes the event horizon is likely to be ``near'' the apparent
horizon. If accretion dominates over evaporation the event horizon is
likely to be just outside the apparent horizon, while if evaporation
dominates over accretion it is likely to be just inside the apparent
horizon. There is an unpopular minority opinion that takes the view
that because of Hawking evaporation a black hole will never form a
true event horizon (absolute horizon). For the derivation presented
herein this entire issue is simply not relevant.]

Define a quantity:
\begin{equation}
g_H(t) 
= {1\over2} \left.{\d  [c(r,t)^2 - v(r,t)^2]\over \d r}\right|_H 
= c_H \left.{\d  [c(r,t)- |v(r,t)|]\over \d r}\right|_H.
\end{equation}
If the geometry is static, this reduces to the ordinary definition of
{surface gravity}~\cite{Visser}:
\begin{equation}
\kappa = {g_H\over c_H}.
\end{equation}
If the geometry is not static this is a natural definition of the
``surface gravity'' of the apparent horizon.

\section{Eikonal and WKB approximation ($s$ wave)}

Consider a quantum field $\phi(r,t)$ on this \Painleve--Gullstrand
background and take the eikonal (geometric optics/acoustics)
approximation for the $s$ wave
\begin{equation}
\phi(r,t)  
= {\cal A}(r,t) \; \exp[\mp i\varphi(r,t)] =   {\cal A}(r,t) \;
{
\exp\left[\mp i\left(\omega\;t - \int^r k(r') \; \d r'\right)\right]}
\end{equation}
whereby the field is written as a rapidly varying phase times a slowly
varying envelope.  The second equality above, where we have written
the time dependence of the phase as $\omega t$, is valid provided the
geometry is slowly evolving on the timescale of the wave, that is,
provided $\omega \gg \hbox{max} \{ |\dot c/c|, |\dot v/v|\} $. (I
take $\omega$ to be positive.)

In the eikonal approximation the d'Alembertian equation of motion becomes
\begin{equation}
g^{\mu\nu} \; \partial_\mu \varphi \; \partial_\nu \varphi +i\epsilon= 0.
\end{equation}
Note in particular that I have used the eikonal approximation to
immediately impose Feynman's ``$i\epsilon$-prescription'' on the field
($\epsilon$ is real, positive, and infinitesimal).
The metric signature is the general relativity standard $-+++$, which
is why the $i\epsilon$-prescription appears reversed relative to the
particle physics standard. Also note that in invoking the prescription
I have implicitly used the fact that the spacetime geometry is smooth,
even at the horizon, so that it makes sense to both adopt an eikonal
approximation and then within this framework use ordinary flat-space
results. The use of the $i\epsilon$ prescription in this way can be
traced back, at least, to the early paper of Damour and
Ruffini~\cite{Damour}. If one prefers, the $i\epsilon$-prescription
can be rephrased in terms of analyticity of the fields on the
complexified past light cone, ${\cal Z}(x) = \{z=x+iy\}$, where $x$ is
any point in spacetime and $y$ lies in the past light cone. This is
the Lorentz-invariant generalization of analyticity in the lower
half-plane, but for all practical purposes can be replaced by the
$i\epsilon$ prescription.

From the preceding equation
\begin{equation}
\omega^2 - 2v(r,t)\; \omega k - [c(r,t)^2-v(r,t)^2] k^2 +i\epsilon= 0.
\end{equation}
So that
\begin{equation}
(\omega - v k)^2 = c^2 k^2+i\epsilon.
\end{equation}
Whence
\begin{equation}
\omega - v k = \sigma\; (1+i\epsilon) \; c k; \qquad \sigma = \pm1.
\end{equation}
For specific real frequency $\omega$ this gives us the wave-vector
$k(r,t)$ as
\begin{equation}
k 
= 
{\omega\over  \sigma\; (1+i\epsilon) \; c + v} 
=
{\sigma \; \omega \over  (1+i\epsilon) \; c + \sigma v} 
= {\sigma \;  (1+i\epsilon) \; c - v\over  (1+i\epsilon)^2\; c^2-v^2} \;\omega.
\end{equation}
Note:
\begin{equation}
\sigma = +1 \qquad \implies \qquad \hbox{outgoing mode}.
\end{equation}
\begin{equation}
\sigma = -1 \qquad \implies \qquad \hbox{ingoing mode}.
\end{equation}
Keeping track of the $i\epsilon$ is important only near the apparent
horizon, where it is critical; everywhere else it can safely be set to
zero.

While we do not really need to use the WKB approximation (physical
acoustics/optics) it can be invoked at very little additional cost.
The (approximate) conserved current is
\begin{equation}
J_\mu = | {\cal A}(r,t) |^2 \;\; ( \omega, k, 0,0 ).
\end{equation}
Then
\begin{equation}
\nabla_\mu \, J^\mu = 0 \qquad \implies \qquad\qquad
| {\cal A}(r,t) | \propto {1\over r}.
\end{equation}
So we can write
\begin{equation}
\phi(r,t)  \approx {{\cal N}\over\sqrt{2\omega}\;r}\;
\exp\left[\mp i\left(\omega\;t - \int^r k(r') \; \d r'\right)\right],
\end{equation}
where ${\cal N}$ is some conveniently chosen normalization, to
be discussed more fully below.

\section{Near-horizon modes: ingoing}

The ingoing wavenumber ($\sigma = -1$) is
\begin{equation}
k_{\mathrm{in}} 
= - {\omega\over  (1+i\epsilon)\; c-v}.
\end{equation}
Thus in the vicinity of the future apparent $r\approx r_H$ (with $v
\approx - c$) the ingoing wavevector is approximately
\begin{equation}
k_{\mathrm{in}}  \to -{\omega\over2c_H}.
\end{equation}
Thus the ingoing modes are approximately
\begin{equation}
\phi(r,t)_{\mathrm{in}} \approx 
{{\cal N}_{\mathrm{in}}\over\sqrt{2\omega} \; r_H} \;
\exp\left[
\mp i \omega \left\{ t + {r-r_H \over 2 c_H } \right\}
\right].
\end{equation}
This means the phase velocity of the ingoing mode as it crosses the
horizon (in coordinate distance per coordinate time) is $2c_H$.
(Phase velocity equals group velocity because there is no dispersion.)
We see that the ingoing modes contain no real surprises and are
relatively uninteresting.

\section{Near-horizon modes: outgoing}

Now consider the outgoing mode $\sigma = +1$
\begin{equation}
k_{\mathrm{out}} 
= 
{\omega\over  (1+i\epsilon) \; c + v}.
\end{equation}
In the vicinity of the future apparent $r\approx r_H$ (with $v
\approx - c$) the outgoing wavevector is approximately
\begin{equation}
k_{\mathrm{out}} 
\approx 
{\omega \over [g_H/c_H] (r-r_H) + i \epsilon \; c_H}.
\end{equation}
But because $\epsilon$ is infinitesimal, and both $g_H$ and $c_H$ are
by hypothesis positive, we can for all practical purposes rewrite this
in terms of the ``principal part'' and a delta function
contribution. That is, near the apparent horizon
\begin{equation}
k_{\mathrm{out}} 
\approx
{c_H \;\omega\over g_H} 
\left\{ \wp\left({1\over r-r_H}\right) - i \pi \; \delta(r-r_H)\right\}.
\end{equation}
Thus we can ignore the $i\epsilon$ unless we are actually intending to
cross the apparent horizon. In particular, just outside the apparent
horizon
\begin{equation}
\int^r k = \int^r {dr' \;\omega\over c(r') - |v(r')|} \approx 
\int^r {dr'\; c_H \;\omega\over g_H (r' - r_H)} 
= { {c_H \;\omega\over g_H} \; \ln[r- r_H]}.
\end{equation}
Therefore (for $r>r_H$)
\begin{eqnarray*}
\phi(r,t)_{\mathrm{out}} &\approx& {\cal N}_{\mathrm{out}}\;
{\exp\left(\pm i \left[\omega c_H\over g_H\right] \;  \ln[r- r_H]\right)
\over\sqrt{2\omega}\;r_H} 
\;\exp\left\{
\mp i \omega t
\right\}
\\
&\approx& 
 {\cal N}_{\mathrm{out}}\;
{{[r- r_H]^{\pm i\omega c_H/g_H}}\over\sqrt{2\omega}\;r_H} 
\;\exp\left\{
\mp i \omega t
\right\}.
\end{eqnarray*}
The phase velocity of the outgoing mode as it crosses the horizon (in
coordinate distance per coordinate time) is zero.

The fact that these outgoing modes have the surface gravity, $\kappa =
g_H/c_H$, show up in such a fundamental and characteristic way is
already strongly suggestive; and this is really all there is to
Hawking radiation.  The logarithmic phase pile-up at the horizon is
characteristic of many derivations of Hawking radiation, in
particular~\cite{Hawking}, and for many readers this will be
sufficient to convince them that Hawking radiation is present under
the current circumstances (slowly evolving apparent horizon without
prejudice as to where the metric comes from).
In fact, this calculation is the easiest and fastest way I know of to
deduce the existence of the phase pileup using completely elementary
methods.

To properly describe the modes that escape to infinity we should
normalize on the half-line $r>r_H$ using the standard Klein-Gordon
norm. This results in replacing ${\cal N}_{\mathrm{out}}$ with some
specific normalization constant ${\cal N}_{\mathrm{escape}}$ whose
precise value we do not need to know.

\section{Straddling the horizon: the Boltzmann factor}
In addition to looking at outgoing modes that escape to infinity, it
is also useful to consider what happens to ``outgoing'' modes that
straddle the apparent horizon.  Inside the apparent horizon the real
part of $k_{\mathrm{out}}$ goes negative, indicating that while the
outgoing mode is trying to escape ``upstream'' it is actually being
overcome by the flow/shift vector and swept back ``downstream''.
Additionally the phase picks up an imaginary contribution, due
ultimately to the $i\epsilon$ prescription,
\begin{eqnarray}
\int^{r^+}_{r_-} k_{\mathrm{out}} &\approx&
\int^{r^+}_{r_-} \d r' \; {c_H \; \omega \over g_H} 
\left\{ \wp\left({1\over r'-r_H}\right) - i \pi \; \delta(r'-r_H)\right\} 
\nonumber\\
&=&
{c_H\; \omega\over g_H} \; 
\left\{ \ln{|r_+ - r_H|\over|r_- - r_H|} - i \pi\right\}.
\end{eqnarray}
So just inside the apparent horizon
%
\begin{equation}
\phi(r,t)_{\mathrm{straddle}(r<r_H)} \approx 
{\cal N}_{\mathrm{straddle}}\;
{{|r- r_H|^{\pm i\omega c_H/g_H}}\over\sqrt{2\omega}\;r_H} \; 
{\exp\left\{ +{\pi\;\omega \;c_H\over g_H}\right\}} \;
\exp\left[
\mp i \omega t
\right].
\end{equation}
Compare with the situation just outside the apparent horizon
\begin{equation}
\phi(r,t)_{\mathrm{straddle}(r>r_H)} \approx 
{\cal N}_{\mathrm{straddle}}\;
{{|r- r_H|^{\pm i\omega c_H/g_H}}\over\sqrt{2\omega}\;r_H} \; 
\exp\left[
\mp i \omega t
\right].
\end{equation}
Note that this straddling mode has associated with it a normalization
factor ${\cal N}_{\mathrm{straddle}}$, which is distinct from
that of the escaping mode ${\cal N}_{\mathrm{escape}}$; the
straddling mode is to be normalized on the entire half-line $r>0$.  In
terms of the Heaviside step function
\begin{eqnarray}
\phi(r,t)_{\mathrm{straddle}} &\approx& 
{\cal N}_{\mathrm{straddle}}\;
\left[ 
\Theta(r_H-r) \; {\exp\left\{ +{\pi\;\omega \;c_H\over g_H}\right\}} +
\Theta(r-r_H)
\right]
\nonumber\\
&& 
\times {{|r- r_H|^{\pm i\omega c_H/g_H}}\over\sqrt{2\omega}\;r_H} \; 
\exp\left[
\mp i \omega t
\right].
\end{eqnarray}
%
See Damour and Ruffini~\cite{Damour}, equation (5b), for an early
occurrence of a very similar statement; see also Massar and
Parentani~\cite{Parentani}, equation (13).
Note note in particular the presence of a {Boltzmann}-like factor
\begin{equation}
 {\exp\left\{ +{\pi\omega c_H\over g_H}\right\}} 
\end{equation}
which (we shall soon see) corresponds to the Hawking temperature
\begin{equation}
k \; T_H = {\hbar \; g_H\over 2\pi\; c_H}.
\end{equation}
For many physicists the presence of this Boltzmann-like factor will be
enough: The occurrence of Boltzmann factors of this type was the key
to the Hartle--Hawking derivation~\cite{Hartle-Hawking}, though they
were working with the full propagator and dealing with past and future
horizons of a maximally extended Kruskal--Szekeres eternal black
hole. Damour and Ruffini~\cite{Damour} demonstrated the existence of
similar Boltzmann factors for mode functions evaluated at the future
Killing horizon of a Kerr--Newman black hole (dispensing with the past
horizon entirely). In the present situation, the same Boltzmann factor
is seen to arise for any slowly evolving apparent horizon.

The imaginary contribution to the integrated wavenumber is, in
slightly disguised form, equivalent to the imaginary contribution to
the action that occurs in the {Parikh}--{Wilczek}
approach~\cite{Wilczek},\footnote{%
Note that {Parikh} and {Wilczek} have subsequently and implicitly made
use of the Einstein equations at the stage when they then relate the
emission process to the entropy change. This comment also applies to
the Massar--Parentani approach.}
and is also related but not identical to the imaginary contribution
arising from complex paths in the approach of {Padmanabhan}
\etal~\cite{Padmanabhan-et-al}.

But it is possible to quite easily do a lot more: While I have so far
carefully not specified specific values for ${\cal
N}_{\mathrm{escape}}$ and ${\cal N}_{\mathrm{straddle}}$, the
relationship between them is very simple. Since the straddling mode is
to be normalized on $(0,+\infty)$, [which we actually approximate by
the full line $(-\infty,+\infty)$], while the escaping mode is only
normalized on the half line $(r_H,\infty)$, we have
\[
|{\cal N}_{\mathrm{straddle}}|^2 
\left[ {\exp\left\{ +{2\pi\;\omega\; c_H\over g_H}\right\}} - 1 \right]
=
|{\cal N}_{\mathrm{escape}}|^2
\]
So the relative normalization is
\[
\left|{ {\cal N}_{\mathrm{straddle}}\over {\cal N}_{\mathrm{escape}}} \right|^2  =
{1\over {\exp\left\{ +{2\pi;\omega\; c_H\over g_H}\right\}} - 1 }.
\] 
It is this Planckian form of the normalization ratio that then leads
to a Planckian distribution for the outgoing flux. (The straddling
mode contains a Planckian distribution of escaping modes.)

Note the physics assumption hidden here: one is assuming that the
quantum vacuum state is that corresponding to
$\phi_{\mathrm{straddle}}$. That is, freely falling observers (who get
to see both sides of the horizon) should not see any peculiarities as
one crosses the horizon.  Picking the quantum vacuum corresponding to
$\phi_{\mathrm{straddle}}$ implies choosing the Unruh vacuum --- and
when we look at this vacuum state far from the horizon we see the
Planckian flux of outgoing particles.

(If for whatever reason you don't like this normalization calculation
or the related thermodynamic arguments you can alternatively use the
phase pileup property directly to perform a {Bogolubov} coefficient
calculation in the style of~\cite{Hawking}; all roads lead to Rome.)

\section{Beyond $s$ wave}

What happens if we go beyond the $s$ wave? There is now some momentum
transverse to the apparent horizon so that
\begin{equation}
\partial_\mu \varphi = (\omega, -k, -k_\perp).
\end{equation}
If we resolve the field in terms of {partial waves}
\begin{equation}
k_\perp^2= {\ell(\ell+1)\over r^2}.
\end{equation}
Then in the {eikonal} approximation (I now suppress the $i\epsilon$,
it has done its job and would now only serve to clutter the formulae)
\begin{equation}
-\omega^2 + 2v(r,t)\; \omega k + [c(r,t)^2-v(r,t)^2] \; k^2 
+  c(r,t)^2 \; k_\perp^2= 0.
\end{equation}
That is
\begin{equation}
(\omega - v k)^2 = c^2 k^2 + c^2  k_\perp^2.
\end{equation}
This is a quadratic for $k$ as a function of $\omega$ and $k_\perp$:
\begin{equation}
k = 
{\sigma\sqrt{c^2\omega^2 - (c^2-v^2)c^2 k_\perp^2} - v \;\omega
\over c^2-v^2}.
\end{equation}
For ingoing modes near the apparent horizon one must evaluate using
{L'H\^opital}'s rule:
\begin{equation}
k_{\mathrm{in}} \to - {\omega^2 - c^2 k_\perp^2\over 2 \; c_H\; \omega}.
\end{equation}
So the ingoing modes {\emph{do}} depend on $k_\perp$.
\begin{equation}
\phi(r,t)_{\mathrm{in}} \approx { {\cal N}_{\mathrm{in}} \over\sqrt{2\omega} 
\; r_H}
\exp\left[
\mp i \omega \left\{ 
t + {(r-r_H)[ \omega^2 - c^2 k_\perp^2]\over 2 \;c_H\; \omega} 
\right\}
\right].
\end{equation}
But that does not matter: The ingoing modes are not the relevant ones.
For the outgoing modes, near the apparent horizon we see
\begin{equation}
k_{\mathrm{out}} \to {c_H\;\omega\over g_H (r - r_H)}.
\end{equation}
That is, for the outgoing modes:
\\
---The near-horizon asymptotic behaviour is {\emph{{independent}}} of $k_\perp$.
\\
---The phase pile-up is {\emph{{independent}}} of $k_\perp$.
\\
---Continuation of the outgoing modes across horizon is
{\emph{{independent}}} of $k_\perp$.
\\
---The {Hawking} temperature  {\emph{{independent}}} of $k_\perp$.
\\
---This behaviour is {universal} for all {partial waves}.
\\
---Adding a mass term corresponds to:
\begin{equation}
c^2 k_\perp^2 \to  c^2 k_\perp^2 + (m \; c^2/\hbar)^2.
\end{equation}
In view of the above we see that the behaviour of the outgoing modes
near the horizon is also completely independent of the mass and
transverse momentum.  Consequently the {\emph{same}} universal Hawking
temperature applies to all masses and all partial waves. (Again, this
is the easiest way I know of to convince oneself by elementary means
that restricting attention to the $s$-wave captures almost all the
essential physics of Hawking radiation.)

Note that while the Hawking temperature is completely independent of
both angular momentum and mass, the grey-body factors are another
matter: they do depend on both angular momentum and mass and are
responsible for effectively cutting off the higher angular momentum
modes.

\section{Essential features}

The only real physics input has been basic quantum physics plus the
existence of a Lorentzian metric with:
\\
--- {an apparent horizon};
\\
--- {non-zero $g_H$};
\\
--- {slow evolution}.
\\
The need for slow evolution of the geometry is hidden back in the
approximation used to write the modes as $\exp(\pm i \omega t)$ times
a position-dependent factor. This makes sense only if the geometry is
quasi-static on the timescale set by $\omega$. So for consistency we
should only trust the Boltzmann factor, and the Planckian nature of
the Hawking radiation, for frequencies greater than $ \hbox{max} \{
|\dot c/c|, |\dot v/v|\}$. In particular, in order for the peak in the
Planck spectrum to be meaningful we require
\begin{equation}
{k T_H\over\hbar} \approx \omega_{\mathrm{peak}}
\gg \hbox{max} \{ |\dot c/c|, |\dot v/v|\}.
\end{equation}
In particular we require
\begin{equation}
{
\left.{\d  [c(r,t)- |v(r,t)|]\over \d r}\right|_H
\gg {\dot c_H\over c_H}
}.
\end{equation}
Near the horizon spatial gradients should dominate over temporal
gradients. In particular the closer the black hole is to extremality
the slower it is permitted to evolve if there is to be any hope for
even a small segment of quasi-thermal spectrum.

That's it. It is truly remarkable how basic and primitive the Hawking
radiation phenomenon is, and how few physical assumptions are really
necessary.

\section{Discussion}

Can the essential conditions for the Hawking effect be further relaxed? 

One obvious question is the use of spherical symmetry, which precludes
direct application of the current approach to Kerr and other rotating
black holes. This is a technical problem, not a fundamental problem,
and working in axisymmetric geometries will be do-able but somewhat
messier. (For Kerr--Newman black holes the Damour--Ruffini analysis
can be adapted to this end~\cite{Damour}.) A tricky point for general
analog model geometries is that without the Einstein equations, and
something like the dominant energy condition, there is no longer any
reason to believe in the zero'th law: the surface gravity and Hawking
temperature can then in principle vary over different parts of the
apparent horizon; these complications were suppressed in the current
article via the simple expedient of enforcing spherical symmetry. (For
Killing horizons there are derivations of the zero'th law that do not
depend on the Einstein equations~\cite{Killing}, but such
considerations lose their force once the horizon becomes time
dependent.)

Secondly, there are simple linguistic issues of definition: How far
can we push the Hawking effect before we should give it another name?
As argued in this article, based on the physics there is a very good
case for keeping the name the same for the effect in arbitrary
``effective geometries'', no matter how derived. Some would even argue
that the Hawking effect and Unruh effects are fundamentally identical;
I prefer to view then as distinct, possibly as two sides of the same
coin --- the response of the quantum vacuum to externally imposed
conditions.

Thirdly: What is the energy source for the Hawking radiation?  The
infalling particles have negative energy as seen from outside the
apparent horizon. In the case of general relativity black holes the
infalling particles serve to reduce the total mass-energy of the
central object, and it is ultimately the total mass-energy of the
black hole that provides the energy emitted in the hawking flux. In
the case of an acoustic black hole the infalling negative energy
phonon steals kinetic energy from the fluid flow used to generate the
acoustic geometry. For effective geometries associated with ``slow
light'' the electromagnetic control field, used to generate EIT
(electromagnetically induced transparency) and so reduce the group
velocity, provides energy to the system which is available to
ultimately be converted to Hawking-like photons. The general message
is that the Hawking effect steals energy from whatever process is used
to set up the effective geometry in question.

Finally, since this point still seems to cause much confusion, I
should make the explicit comment that
\begin{itemize}
\item {\sl Hawking radiation is not a test of quantum gravity.}
\end{itemize}
Instead, searching for Hawking radiation is a test of the general
principles of quantum field theory in curved spacetimes. As such it is
an ingredient useful for testing semiclassical quantum gravity, though
it does not necessarily probe quantum gravity itself. In particular,
all the proposed experimental tests of Hawking radiation via ``analog
models'' will only probe kinematic aspects of black hole physics.  To
start to address the dynamics of general relativity black holes one
needs the Einstein equations (or some approximation thereto), in which
case one can begin to discuss Bekenstein entropy (or some
approximation thereto). This requires a whole extra layer of
theoretical superstructure, and a key point of this paper is that it
is important and useful to keep these notions logically distinct.

\acknowledgments

I acknowledge comments and feedback from the participants at the Black
Holes III conference in Kananaskis, Canada. I particularly wish to
thank Bei-Lok Hu, Ted Jacobson, Renaud Parentani, and Amanda Peet for
their comments.


\end{document}